# Nonlinear Fiber System for Shot-noise Limited Intensity Noise Suppression and Amplification

**MARVIN EDELMANN,**[1,2,3] **YI HUA,**[1,4] **KEMAL ŞAFAK,**[3] **AND FRANZ X. KÄRTNER**[1,4,*]

[1] *Center for Free-Electron Laser Science (CFEL), DESY, Notkestr. 85, 22607 Hamburg, Germany*
[2] *Department of Physics, Universität Oldenburg, Ammerländer Heerstr. 114-118, 26111 Oldenburg, Germany*
[3] *Cycle GmbH, Notkestr. 85, 22607 Hamburg, Germany*
[4] *Department of Physics, Universität Hamburg, Jungiusstr. 9, 20355 Hamburg, Germany*
*Corresponding author: franz.kaertner@desy.de*

**We propose a nonlinear fiber system for shot-noise limited, all-optical intensity-noise reduction and signal amplification. The mechanism is based on the accumulation of different nonlinear phase shifts between orthogonal polarization modes in a polarization-maintaining fiber amplifier in combination with an implemented sinusoidal transmission-function. The resulting correlation between the input intensity-fluctuations and the system transmission enables tunable intensity noise reduction of the input pulse train. In the experiment, the noise spectral density of a mode-locked oscillator is suppressed by up to ~20 dB to the theoretical shot-noise limit of the measurement at -151.3 dBc/Hz with simultaneous pulse amplification of 13.5dB.**

Intensity fluctuations of optical pulse trains generated by mode-locked lasers are a limiting factor for a large variety of scientific fields and technological applications such as frequency metrology [1], synchronization and timing [2], arbitrary waveform generation [3] and optical microwave extraction [4]. Due to the fast progress in the above-mentioned fields, the demand for systems that are able to generate low-noise pulse trains becomes increasingly important. For this reason, compact fiber-optic laser systems consisting of polarization-maintaining (PM) fibers have established themselves as promising sources for low-noise applications during the last few years [5-7]. While fiber oscillators mode-locked with a nonlinear amplifying/optical loop mirror (NOLM/NALM) already demonstrated sub-fs timing jitter [8] and integrated relative intensity noise (RIN) levels comparable to commonly used solid-state laser systems [9,10], there are still certain inherent limitations on the noise performance of these oscillators that have to be resolved. It is well-known, that the high intracavity loss associated with often required high output coupling ratios from fiber lasers results in enhanced amplified spontaneous emission (ASE) that directly couples to the timing-jitter and intensity noise [11,12]. In addition, nonlinear effects in fibers limit the intracavity pulse energy and allow for a strong conversion of intensity to timing fluctuations based on self-steepening and spectrally limited optical gain [13]. In addition, unbalanced net cavity dispersion directly converts center frequency fluctuations into timing fluctuations [14], a phenomenon known as Gordon-Haus effect [15]. To a certain extent, the influence of these effects can be minimized for example through precise dispersion-engineering [16], by optimizing the cavity loss [12] or by utilizing inherent noise suppressing mechanisms provided by the saturable absorber [17]. Further reduction of the noise far beyond those limitations requires modifications of the cavity, e.g., with an implementation of tailored spectral reshaping with bandpass filters [18,19] or the application of electronic feedback loops for a modulation of the pump source [20]. However, in the first case one has to inevitably trade-off spectral bandwidth and output power from the oscillator which might be crucial parameters for certain applications; whereas in the case of feedback loops, the noise suppression is inherently bandwidth-limited by the electronic circuit while resonant peaks in the noise spectrum often lead to a degradation of the RIN [21].

In this letter, we propose an all-optical fiber amplifier system for highly efficient and broadband intensity noise suppression of given input pulse trains. The underlaying mechanism is based on the nonlinear co-propagation of degenerate polarization-modes in a PM-fiber amplifier and the resulting accumulation of a nonlinear phase difference $\Delta\varphi_{nl}$ between both modes. The intensity-dependence of this process transforms input intensity-fluctuations into proportional fluctuations of $\Delta\varphi_{nl}$. An additional implementation of a sinusoidal transmission function $T(\Delta\varphi_{nl})$ induces a periodic intensity-noise transfer function, tunable through a non-reciprocal phase-bias. Simulations and experimental results give insight into the possibility to tailor the system parameters for most effective intensity-noise suppression with maximal signal amplification. Systematic measurements reveal that certain sets of parameters enable shot-noise limited suppression of the input noise spectral density over a broad bandwidth by up to 20 dB with simultaneous signal amplification of >13dB.

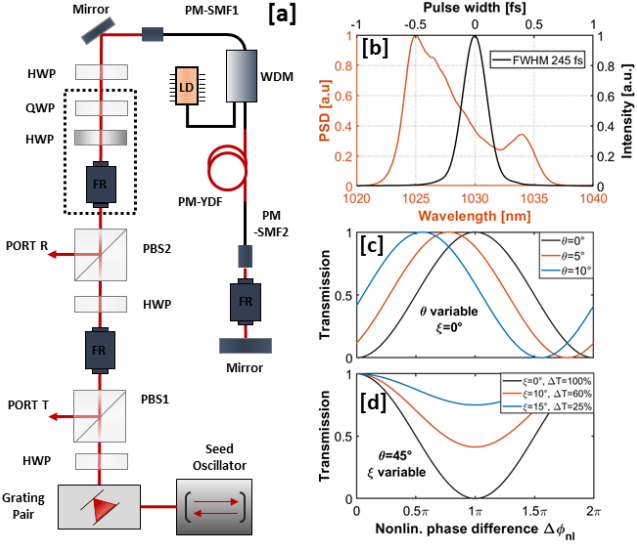

**Fig.1:** [a] Schematic of the nonlinear all-PM fiber amplifier system for intensity noise suppression. PBS: Polarization beam-splitter, FR: Faraday-rotator, HWP: Half-wave plate, QWP: Quarter-wave plate, SMF: Single-mode fiber, YDF: Ytterbium-doped fiber. [b] Measured spectral bandwidth (orange) and autocorrelation trace (black) of the seed oscillator with a FWHM of 5 nm and 245 fs, respectively. [c]: Transmission function $T(\Delta\varphi_{nl})$ of the system for different HWP rotation angles θ with the QWP rotation angle ξ fixed at 0°. [d]: Nonlinear transmission $T(\Delta\varphi_{nl})$ for a variable ξ and θ fixed at 45°.

A schematic of the experimental setup is shown in Fig.1[a]. The seed oscillator is an Ytterbium (Yb)-doped fiber oscillator mode-locked by nonlinear polarization evolution (NPE) in the all-normal dispersion regime. The repetition rate is ~40 MHz with an average output power of 30 mW. The measured output spectrum and the autocorrelation trace of the compressed pulse are shown in Fig.1[b] with a 3 dB bandwidth of 5 nm and a pulse duration of 245 fs, respectively. The output pulses are compressed with a 1000 lines/mm transmission grating pair. The non-reciprocal phase bias consists of a 45°-Faraday-rotator (FR), a half-wave plate (HWP) and a quarter-wave plate (QWP). An additional HWP aligns the slow axis of the PM-fiber segment to the transmission axis of PBS. The fiber segment consists of 1 m single-mode polarization maintaining fiber (SM-PMF), 0.4 m PM-YDF and another 2 m SM-PMF in a sequence. The YDF is optically pumped through a WDM with a 1 W, 976 nm laser diode. After a single pass through the PM-fiber amplifier, a Faraday-mirror (FM) rotates the polarization modes by 90° and reflects them back, compensating the drift between both modes caused by the different group and phase velocities in the birefringent fiber.

As a consequence of the non-reciprocal phase bias and the nonlinear phase difference $\Delta\varphi_{nl}$ accumulated between the polarization modes, the described system in Fig.1[a] creates a tunable sinusoidal transmission function $T(\Delta\varphi_{nl})$ as it is also present e.g., in NALM mode-locked fiber lasers. Hence, the conversion of peak power fluctuations $P=\overline{P}+\delta P(t)$ into fluctuations of the nonlinear phase difference $\Delta\varphi_{nl}=\overline{\Delta\varphi_{nl}}+\Delta\varphi_{nl}(t)$ couples the system transmission with the input intensity fluctuations. For the case of a negative derivative at $T(\overline{\Delta\varphi_{nl}})$ and a positive sign of $\overline{\Delta\varphi_{nl}}$, the transmission response of the system results in an effective suppression of the input noise [17]. Depending on the energy splitting ratio ε, even vanishingly low input fluctuations δP(t) can be converted to large corresponding values of $\Delta\varphi_{nl}(t)$ inducing a strong effect of the system transmission response. The shape of $T(\Delta\varphi_{nl})$ and the energy splitting ratio ε between the degenerate modes depend on the QWP (ξ) and HWP (θ) rotation angles in the phase-bias. Subsequently, all rotation angles are measured with respect of the PBS1 transmission axis. Fig.1[c] shows the transmission $T(\Delta\varphi_{nl})$ for a fixed QWP angle ξ=0° and varying HWP angle θ. This specific setting allows a controlled shift of $T(\Delta\varphi_{nl})$ with simultaneously fixed values of the modulation depth at ΔT=100% and $\overline{\Delta\varphi_{nl}}=0$. The derivative of $T(\Delta\varphi_{nl})$ reaches its maximum value for this state while the energy splitting ratio ε always remains at a constant value of 0.5. Another degree of freedom for $T(\Delta\varphi_{nl})$ is shown in Fig.1[d] with θ=45° and a variable QWP rotation angle ξ. This setting enables the possibility to adjust the energy splitting ratio **ε** and the modulation depth ΔT for a fixed phase bias between the degenerate polarization modes. In addition to the phase bias, the gain in the Ytterbium-doped fiber (YDF) can be utilized for a controlled tunability of $\Delta\varphi_{nl}$, while simultaneously amplifying the input pulse energy.

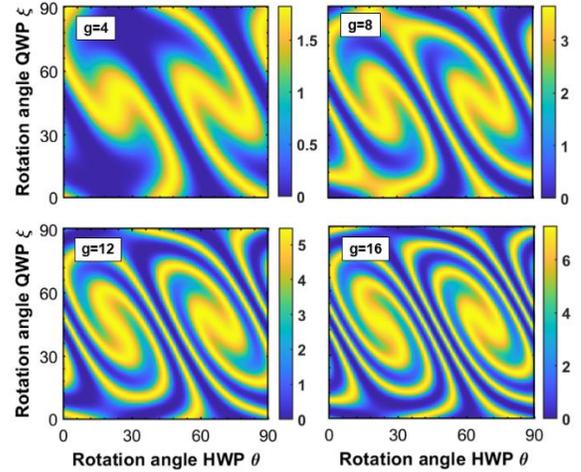

**Fig.2:** Simulated intensity-noise transfer factor RIN$_{out}$/RIN$_{in}$ at port T as a function of the QWP and HWP rotation angles for different gain factors in the YDF. The input pulse train in the simulation has an average peak power of $\overline{P}$=1kW with ±1% fluctuations.

Fig.2 shows the simulated noise-transfer coefficient RIN$_{out}$/RIN$_{in}$ as a function of the phase bias rotation angles ξ and θ for an increasing gain factor in the YDF from g=4 to g=16. The simulation is based on the model described in Ref. [17] with an input average peak power of $\overline{P}$=1kW and a fluctuation δP of ±1%. The fiber segment consists of a 2-m long fiber with the parameters of standard 1 µm PM-fibers (Nufern PM980-XP). As it can be seen, the coefficient RIN$_{out}$/RIN$_{in}$ can be increased or decreased based on the values of ξ and θ with RIN$_{out}$/RIN$_{in}$=0 corresponding to a complete suppression of the input intensity noise. Higher gain in the YDF increases the sensitivity of the noise transfer with respect to changes in the phase bias rotation angles. In addition, the modulation depth of the noise transfer increases proportional to the gain, which allows for stronger modulation of the input noise RIN$_{in}$ in both directions for higher gain factors. Another remarkable aspect of the fiber system is the amplified pulse energy at the output port while simultaneously suppressing the noise. A measurement of the noise transfer coefficient RIN$_{out}$/RIN$_{in}$ at the output Port T together with the achievable gain can be seen in Fig.3. Here, the noise transfer is shown together with the measured gain as a function of the HWP angle $\theta$ for fixed QWP angles $\xi$ at 30° and 60°, respectively.

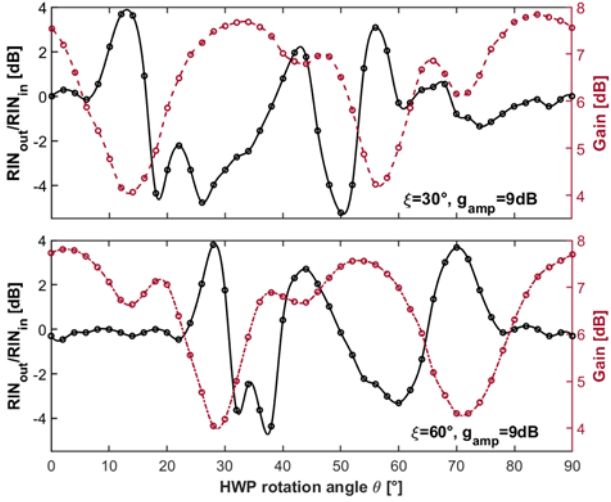

**Fig.3:** Noise transfer $RIN_{out}/RIN_{in}$ (black) and corresponding gain (red) measured at Port T as a function of the HWP rotation angle $\theta$ for fixed QWP angle $\xi$ =30° (top) and $\xi$ =60° (bottom). The total gain in the YDF is 9 dB and the RIN value is integrated from 1kHz to 10 MHz.

The total optical gain in the YDF is 9 dB which is then distributed between Port R and Port T based on $T(\Delta\varphi_{nl})$. In order to determine the noise transfer curves, the RIN is measured with the AM-noise function of a signal-source analyzer (SSA, Keysight E5052B). Therefore, the pulse train is measured with an InGaAs-photodetector (EOT ET-3000) and the second harmonic of the received RF signal is filtered out. Subsequently, a low-noise 13 dBm RF-amplifier (Mini-Circuits ZX60-33LN-S+) increases the RF-power that finally reaches the SSA to -2 dBm. The RIN value is obtained via integration of the AM-noise spectrum in the range of 1 kHz to 5 MHz. As can be seen in Fig.3, the maximum suppression of the input RIN for both states of $T(\Delta\varphi_{nl})$ is ~5.7 dB from $RIN_{in}$=0.037% to $RIN_{out}$=0.010% with a simultaneous gain of ~7 dB in both cases. The working points with highest efficient noise suppression correlate with local peaks of the corresponding gain curves at Port T. In addition, a comparison of the noise transfer and gain curves in Fig.3 for $\xi$ =30° and $\xi$ =60° show a periodic behavior with respect to the HWP rotation angle in agreement with the simulations shown in Fig.2.

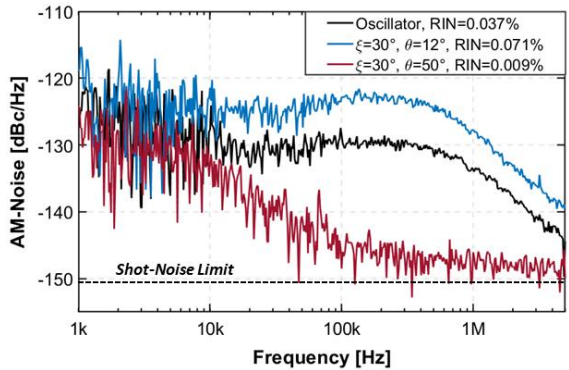

**Fig.4:** AM-Noise spectral density of the oscillator in the range from 1kHz to 5 MHz (black, RIN=0.037%) compared with the state of the system where the RIN is utmost suppressed by 5.6 dB (red, RIN=0.009%) and deteriorated by ~4 dB (blue, RIN=0.071%).

Fig.4 shows the measured frequency-resolved AM-noise spectral density of the auxiliary oscillator compared to the spectra measured at Port T with utmost suppressed and deteriorated RIN in Fig.3 ($\xi$=30°) by -5.6 dB and 4 dB, respectively. While the low frequency noise is equal for all three cases due to the identical environment of the measurements, the maximum noise-suppressed state of the system with $\xi$=30° and $\theta$=50° ensures a reduction of the noise spectral density by up to 17.5 dB for the frequency range from 100 kHz to 1 MHz. Since the noise transfer is purely based on the non-resonant optical Kerr-effect, the system is able to respond almost instantaneously to the fast intensity-fluctuations as long as they are above fundamental limitations. For higher frequencies (>1MHz) the noise difference decreases as the oscillator and the noise amplified state ($\xi$=30°, $\theta$=12°) are approaching the shot noise limit at -151.3 dBc/Hz. Here, the shot noise is calculated as $S_{sn}(f) = 2h\upsilon/\overline{P}$ with the measured RF-signal mean voltage of 37 mV, a 100 $\Omega$ termination and a 0.75 W/A responsivity of the detector at 1030 nm [22]. Shot-noise limitation for the measurements is assumed due to the required attenuation of the amplified optical output signal prior to photodetection [23]. The noise floor of the noise suppressed state in Fig.4 reaches ~-148 dBc/Hz. In order to demonstrate truly shot-noise limited noise suppression with simultaneous decent amplification, another experimental approach is carried out which includes the step-wise shift of $\overline{\Delta\varphi_{nl}}$ through the gain in the YDF with a fixed state of $T(\Delta\varphi_{nl})$.

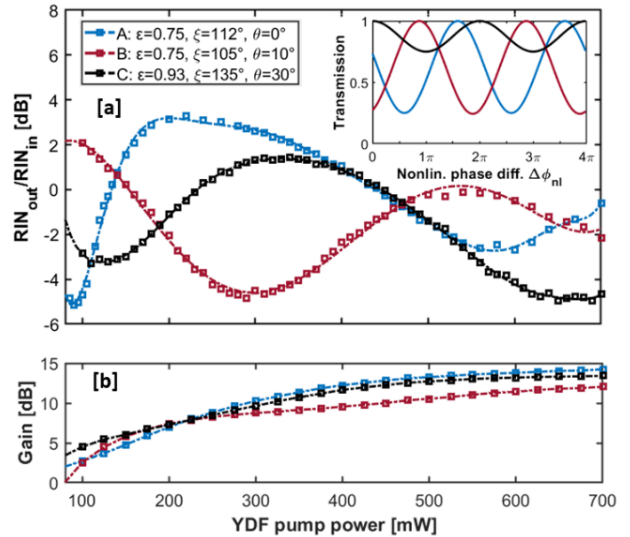

**Fig.5:** [a] Noise transfer coefficient $RIN_{out}/RIN_{in}$ measured at Port T with respect to YDF pump power for different working points of the system A, B and C (blue, red and black), corresponding to the transmission functions shown in the inset. The resulting energy splitting ratios between the polarization modes are $\varepsilon$=0.75 (A, B) and $\varepsilon$=0.93 (C), respectively. [b] Gain at Port T for each working point with respect to an input pulse energy of 0.25 nJ.

As can be seen in Fig.5, a shift of $T(\Delta\varphi_{nl})$ with fixed energy splitting ratio $\varepsilon$=0.75 for states A and B enables a proportional shift of the noise transfer curves, with respect to pump power in the YDF and the corresponding gain. For both states, the maximum suppression of the input RIN by 5dB (A) and 4.8dB (B) correspond to a simultaneous amplification of the input pulse energy by ~3dB and 8.7dB, respectively. For an energy splitting ratio of $\varepsilon=0.93$ as it is present in state C, the nonlinear system enables similar effective suppression of the RIN by 5 dB for an even higher gain up to 13.5dB at port T.

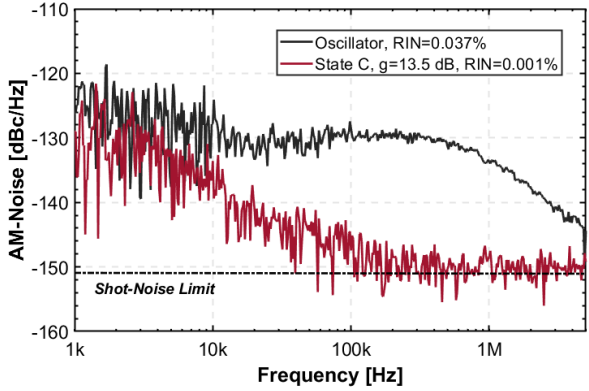

**Fig.6:** AM-noise spectral density in the range from 1 kHz to 5MHz for state C of the nonlinear amplifier (red) in Fig.5 and the input noise spectral density of the oscillator (black). The corresponding RIN suppression is 5.6 dB with a simultaneous signal amplification of 13.6 dB.

Fig. 6 shows the AM-noise spectral density for working point C with the utmost RIN suppression of 5dB at an YDF pump power of 640 mW in comparison with the oscillator noise spectrum. Here, the system suppresses the input RIN of 0.037% ($RIN_{in}$) by 5.6dB down to 0.01% ($RIN_{out}$) measurable at Port T. Simultaneously, the input pulse energy is amplified by 13.5 dB from 0.25 nJ to ~2 nJ. The AM-noise spectral density is reduced by up to 20 dB for the frequency range around 500 kHz. Above 100 kHz, the suppression of the AM-noise reaches the calculated shot noise limit of -151.3 dBc/Hz. For frequencies <10 kHz, the noise spectra of the oscillator and the nonlinear amplifier are again almost identical due to environmental perturbations and a similar setup for the measurement. In the experiment, a comparable effective noise suppression can be achieved for arbitrary YDF pump powers (signal gain), once the setting of the phase bias rotation angles $\xi$ and $\theta$ ensure a fitting energy splitting ratio $\varepsilon$ with a positive value of $\overline{\Delta\varphi_{nl}}$ and a negative derivative of $T(\overline{\Delta\varphi_{nl}})$.

In conclusion, we developed and investigated a nonlinear fiber system which is able to suppress intensity fluctuations of pulsed input signals down to the shot-noise limit over a broad offset-frequency bandwidth. The mechanism is based on the accumulation of a nonlinear phase difference $\Delta\varphi_{nl}$ between orthogonal polarization modes in a standard PM-fiber with compensation of linear phase shifts and the subsequent interaction of the polarization-modulated pulse with an artificial sinusoidal transmission function $T(\Delta\varphi_{nl})$. The high degree of tunability in the phase bias and the YDF gain show the possibility for tailored noise suppression matched to the characteristics of the respective input pulse train. In the experiment, a suppression of the RIN by up to 5.6 dB with simultaneous amplification by 13.5 dB has been demonstrated with the noise spectral density being suppressed by up to 20 dB down to the shot-noise limit of -151.3 dBc/Hz for the frequency range >100kHz. The development of an all-optical system for shot-noise limited intensity noise suppression of pulsed input signals with simultaneous amplification of the pulse energy is another step towards ultralow-noise ultrafast fiber laser systems. Since the basic physical mechanism does not depend on the amplification regime or the fiber-type, we believe that it can be well suited for pulsed high-power amplification with extra-cavity noise suppression.

**Disclosures.** The authors declare no conflicts of interest.